\documentclass[journal]{IEEEtran}
\usepackage{amsmath}
\usepackage{mathtools}
\usepackage{amssymb}
\usepackage{amsfonts}
\usepackage{graphicx}
\usepackage{multirow}
\usepackage{float}
\usepackage{wrapfig}
\usepackage{cite}
\usepackage{capt-of}
\usepackage[justification=centering, font=footnotesize]{caption}
\usepackage{url}
\usepackage[export]{adjustbox}
\usepackage[caption = false]{subfig}
\makeatletter
\def\@IEEEsectpunct{\ \,}
\makeatother
\IEEEoverridecommandlockouts
\ifCLASSINFOpdf
\else
\fi

\begin{document}

\title{Analysis of Handover Failures in Heterogeneous Networks with Fading}

\author{\IEEEauthorblockN{
Karthik Vasudeva\IEEEauthorrefmark{1},
Meryem \c{S}imsek\IEEEauthorrefmark{2},
David L\'opez-P\'erez\IEEEauthorrefmark{3} and
\.{I}smail G\"{u}ven\c{c}\IEEEauthorrefmark{1}}\\
\IEEEauthorblockA{\IEEEauthorrefmark{1}Electrical and Computer Engineering, Florida International University, USA\\
\IEEEauthorrefmark{2}Dresden University of Technology, Germany\\
\IEEEauthorrefmark{3} Bell Laboratories, Alcatel-Lucent, Ireland\\
Email: {\tt \{kvasu001, iguvenc\}@fiu.edu}, {\tt meryem.simsek@tu-dresden.de}, {\tt dr.david.lopez@ieee.org}
}}

\maketitle

\vspace{-9ex}
\begin{abstract}
The handover process is one of the most critical functions in a cellular network, 
and is in charge of maintaining seamless connectivity of user equipments (UEs) across multiple cells. 
It is usually based on signal measurements from the neighboring base stations (BSs),
and it is adversely affected by the time and frequency selectivity of the radio propagation channel. 
In this paper, we introduce a new model for analyzing handover performance in heterogeneous networks (HetNets) 
as a function of vehicular user velocity, cell size, and mobility management parameters. 
In order to investigate the impact of shadowing and fading on handover performance, 
we extract relevant statistics obtained from a 3rd Generation Partnership Project (3GPP)-compliant HetNet simulator,
and subsequently, we integrate these statistics into our analytical model to analyze handover failure probability under fluctuating channel conditions. 
Computer simulations validate the analytical findings, 
which show that fading can significantly degrade the handover performance in HetNets with vehicular users.
\end{abstract}


\IEEEpeerreviewmaketitle

\section{Introduction}
%
%
%
%
The new generation of wireless user equipments (UEs) have made user data traffic and network load to increase in an exponential manner, 
straining current cellular networks to a breaking point \cite{cisco_report}. 
Heterogeneous networks (HetNets) which consist of traditional macrocells overlaid with small cells (e.g., picocells, femtocells, phantom cells etc.), 
have shown to be a promising solution to cope with this wireless capacity crunch problem \cite{LTE_vision2020}. 
Due to their promising characteristics, 
HetNets have gained much momentum in the wireless industry and research community during the past several years. 
For instance, there have been dedicated study and work items in the third generation partnership project (3GPP) related to HetNet deployments~\cite{3GPP_study_items}. 
Their evolutions are also one of the major technology components  that are being considered for 5G wireless systems~\cite{Andrews_5G_2014}.

Despite their promising features, 
HetNets have introduced new challenges, such as the mobility management. Handover among different base stations (BSs) is the main process that supports seamless connectivity of UEs to the network. 
Due to increased number of cells in the network, 
it is difficult to support seamless mobility of UEs in a HetNet scenario
since handovers may fail~\cite{HetNet_mob_challenges}. 
In particular, using the same set of handover parameters of a traditional macrocellular network for a HetNet scenario will degrade the mobility performance of the UEs~\cite{David_mag_2012}. 
For handover in Long Term Evolution (LTE) systems, signal measurements obtained at a UE from the neighboring BSs are reported by the UE to its serving BS, 
and the handover decision is made by the serving BS. 
In LTE HetNets, 
due to the small cell sizes, such measurement reporting by the UE may not be finalized sufficiently quickly, and this might result in severe handover failure (HF) problems for the high velocity users~\cite{TR36839}.




Handover performance has been studied for homogeneous~\cite{Dimou2009, Jansen2011,Zhang2012,Munoz2013,Aziz2010,Anas_PIMRC_2007,Aziz_VTC_2009} and heterogeneous~\cite{Gao2013,Barbera2012,Gelabert2013,David_WCNC_2012,Barbera_Globecom_2012,Barbera2013,Kim2011,Mehbodniya2013,Peng2012} network deployments in the literature. 
In homogeneous networks, 
the authors in~\cite{Dimou2009} use computer simulations to investigate the handover performance of LTE networks, considering different measurement filtering parameters at the UE. 
A novel self-organizing handover management technique is proposed in~\cite{Zhang2012,Munoz2013,Jansen2011,Aziz2010}, where the network autonomously configures the mobility management parameters for different scenarios, 
thereby improving the handover performance of the homogeneous cellular network. Handover parameters (e.g. time-to-trigger (TTT),  hysteresis threshold, etc.) are optimized in~\cite{Gao2013} to achieve robust and seamless mobility of UEs in a \emph{HetNet} scenario. In~\cite{Barbera2012}, mobility performance of UEs is evaluated in the co-channel small cell networks scenario; when the density of the small cell increases, switching off the macro cell is shown to provide seamless mobility for the low speed UEs, while it degrades the handover performance for the high speed UEs~\cite{Gelabert2013}. Furthermore, in~\cite{David_WCNC_2012} authors show that using intercell interference coordination (ICIC) techniques can enhance the handover performance for both low and high speed UEs. Mobility state estimation is performed in~\cite{Barbera_Globecom_2012} to estimate the velocity of the UEs and thereby keeping the high speed UEs to macrocells and low speed UEs are offloaded to picocells, thereby enhancing the handover performance of the UEs. In~\cite{Barbera2013}, mobility performance is analyzed with and without inter-site carrier aggregation for macro and pico cells deployed on a different carrier frequencies. The authors in~\cite{Kim2011,Mehbodniya2013} aim to improve the mobility performance of UEs across different network types such as WiFi, WiMAX, LTE, and Bluetooth, by performing a vertical handoff (VHO). 

Despite all these related work on mobility management in HetNets, there are only limited theoretical studies that analyze the handover performance in HetNet scenarios. 
In~\cite{Andrews_Mobility_2012}, the authors derive the handover rate and sojourn time of a UE for the Poisson-voronoi and hexagon cellular topologies. 
Expressions for call block and drop probabilities in a small cell scenario are derived in~\cite{Sreenath_IOFC_2010}. 
Theoretical analysis for handover performance optimization is done in~\cite{Guidolin2014} to quantify the user performance as a function of user mobility parameters. 
In~\cite{Nguyen2011}, a mathematical framework was proposed to model the handover measurement function, and expressions were derived for measurement failure and best target cell. In~\cite{David_ICC_2012}, handover performance analysis was performed as a function of handover parameters, 
e.g. TTT and UE velocity, 
and in~\cite{Vasudeva2014} the analysis was extended to consider layer-3 measurement filtering process at the UE. To the best of our knowledge, apart from our preliminary results in~\cite{Vasudeva_ICC_2015}, there are no analytical results in the literature that study the HF probability in HetNets as a function of different mobility management parameters and under fading channel conditions.

The main goal of this paper is to introduce a simple yet effective model for analyzing handover failures in small cell deployments, considering all important mobility management parameters of interest. Handover trigger locations at a picocell, and radio link failure locations at a macrocell and a picocell are modeled using co-centric circles. Considering a linear mobility model for UEs, HF probabilities for macrocell and picocell UEs are derived in closed form for various scenarios. The analysis is then extended to fast-fading and shadowing scenarios: relevant statistics in a fading scenario are extracted from a 3GPP compliant system level simulator, to facilitate semi-analytic expressions for handover failure probabilities. All theoretical results are validated through simulations, where impact of different parameters on handover failure and ping-pong probabilities are investigated.

The paper is organized as follows. 
In Section~\ref{Sec: handover_process_lte}, the handover process in LTE and the handover measurement process in a UE are reviewed. 
In Section~\ref{Sec: analytical_model}, a new analytical model for handover performance analysis is presented. 
In Section~\ref{Sec: HF_analysis}, HF probability expressions are derived in the absence of fading and shadowing, while a semi analytical approach for analyzing the handover performance in fast fading environments is introduced in Section~\ref{Sec: HF_analysis_fading}.  
In Section~\ref{Sec: simulation_results}, the theoretical expressions for the HF are verified via simulations, and the last section provides the concluding remarks.

\section{Review of the Handover Process in LTE} \label{Sec: handover_process_lte}
\subsection{Different Stages of Handover Process}

\begin{figure}[h]
\begin{center}
\includegraphics[width=3.5in]{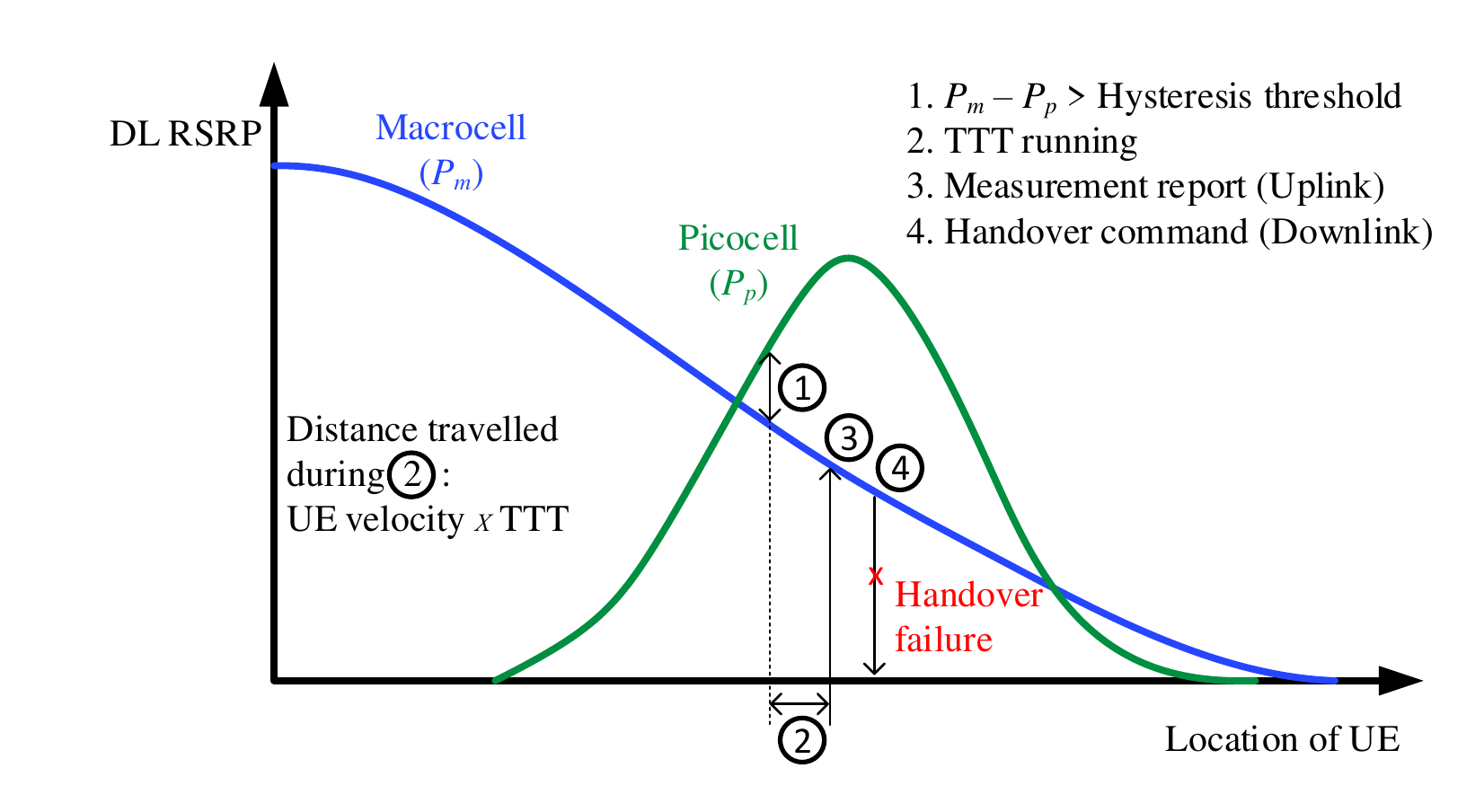}
\captionof{figure}{Handover failure problem in HetNets due to small cell size.}
\label{Fig:HF_HetNets}
\end{center}
\end{figure}

The key steps of a typical handover process in a HetNet scenario are illustrated in Fig.~\ref{Fig:HF_HetNets}. 
Handover decisions are based on the signal strength measurements of the neighboring BSs done at the UE. 
In LTE, UEs perform reference signal received power (RSRP) measurements to assess the proximity of neighbouring cells. 
An example for the downlink (DL) RSRP measurement profile of a macrocell and a picocell, measured by a mobile UE, are shown in Fig.~\ref{Fig:HF_HetNets}. 
Once the measurements are performed,
the UE checks for the handover event entry condition, e.g., 
when the signal strength $P_{\rm p}$ from target cell (e.g., a picocell) is larger than the signal strength from a serving cell (e.g., a macrocell) $P_{\rm m}$ plus a hysteresis threshold (step-1). 
Even when this condition is satisfied, 
the UE waits for a duration of TTT, before sending a measurement report to its serving cell (step-2).
 
The use of a TTT is critical to ensure that ping-pongs (successive and unnecessary handovers among neighboring cells), 
generated due to fluctuations in the link qualities from different cells, 
are minimized. 
If the  handover event entry condition is still satisfied after the TTT, 
the UE sends the measurement report to its serving BS in its uplink (step-3), 
which then communicates with the target cell.
If both cells have an agreement and the handover is to be performed, 
the serving BS  sends a handover command to the UE in the DL to indicate that it is should connect to the target cell (step-4).
The handover process is finalized when the UE sends a handover complete to the target cell,
indicating that the handover process was completed successfully (step-5).

\vspace{-3mm}
\subsection{Handover Measurements Procedure in LTE.}\label{Subsec: handover_process_lte}

\begin{figure}[h]
\begin{center}
\subfloat[Handover measurement model specified in \cite{TS25302}.]{\includegraphics[width=3.5in]{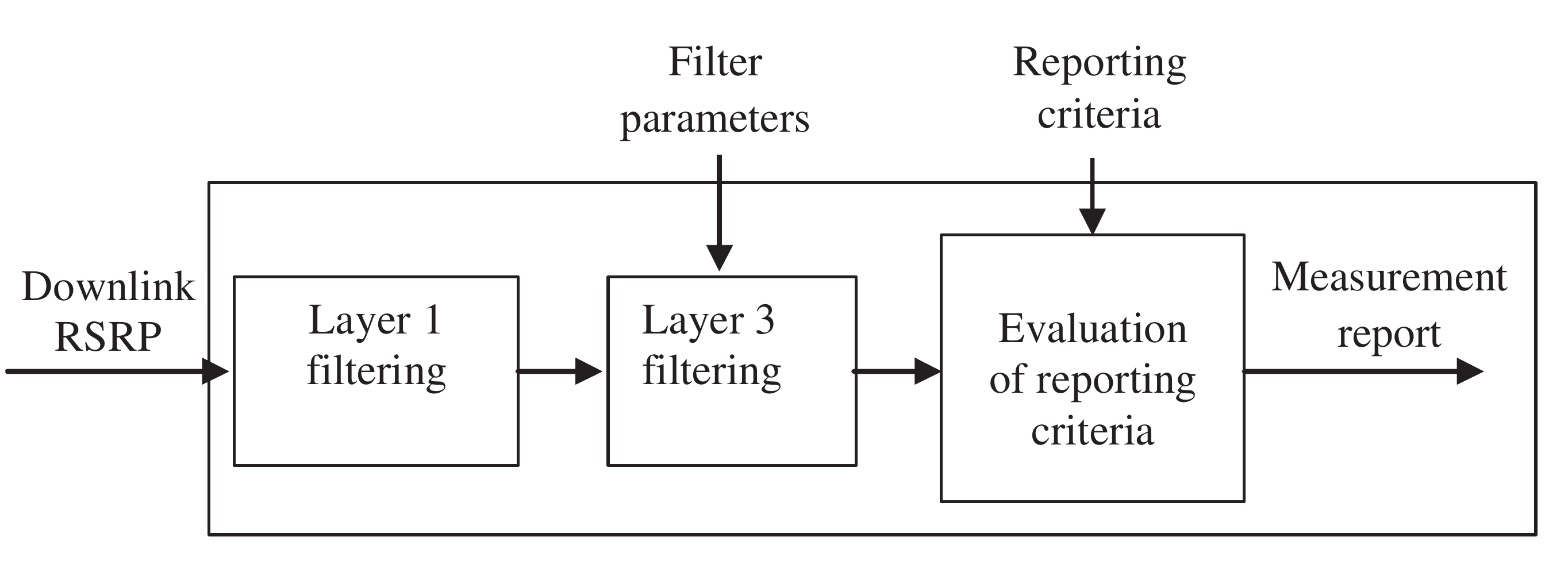}
\label{Fig:HO_measurement_model_3GPP}}\\
\subfloat[Processing of the RSRP measurements through L1 and L3  filtering at a UE ~\cite{David_mag_2012}.]{\includegraphics[width=3.5in]{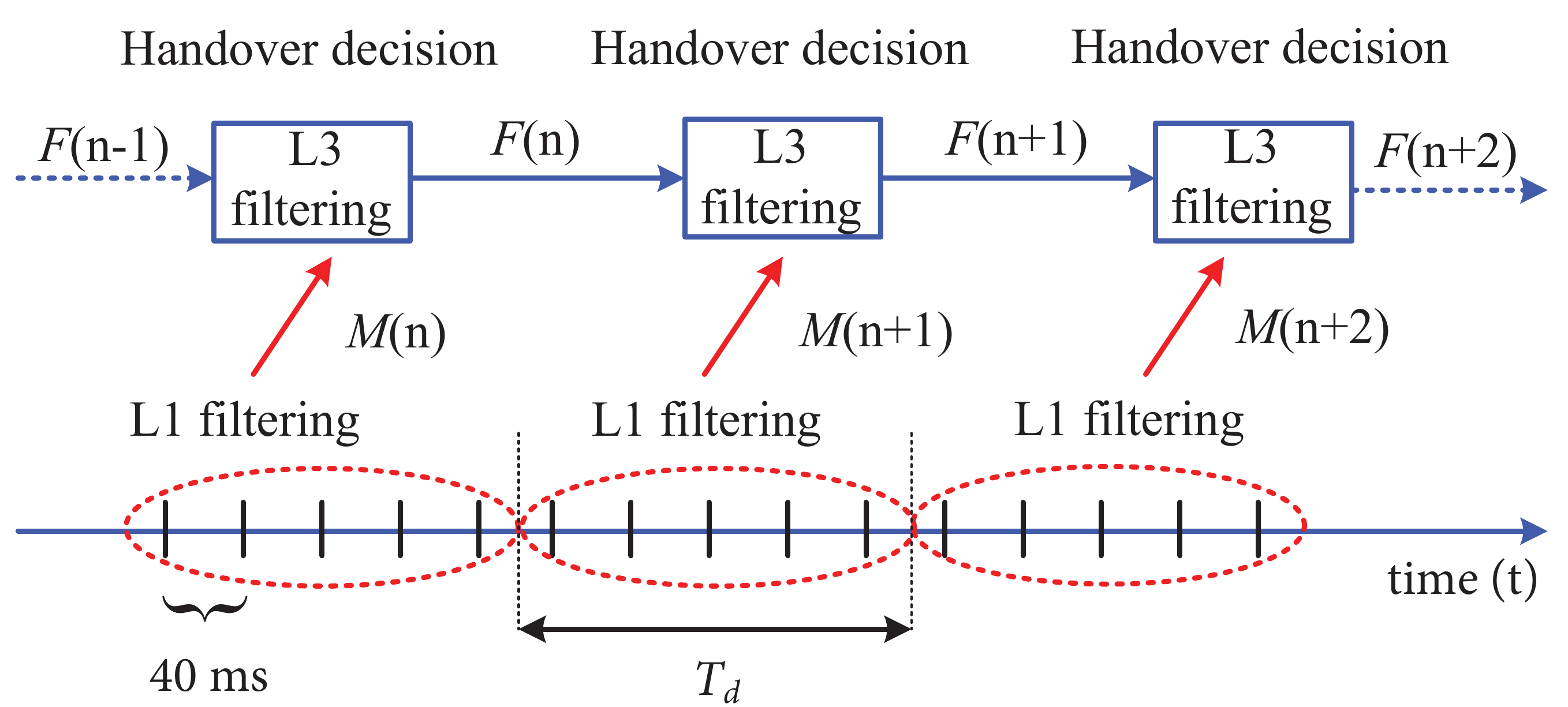}
\label{Fig:HO_measurement_model_David}}
\captionof{figure}{Handover measurement performed by the UE through filtering process in LTE.}
\label{Fig:HO_measurement_model}
\end{center}
\end{figure}


Different measurements obtained at a UE during a handover process are summarized in Fig.~\ref{Fig:HO_measurement_model}. 
It is important to note that the RSRP measurements $P_{\rm m}$ and $P_{\rm p}$ at a UE are obtained after a filtering process for smoothing the signals, in order to mitigate the effects of channel fluctuations.
The filtering of handover measurement is performed in Layer-1 (L1) and Layer-3 (L3), 
as shown in Fig.~\ref{Fig:HO_measurement_model}\subref{Fig:HO_measurement_model_3GPP}. 
Initially, the UE obtains each RSRP sample by linear averaging over the power contribution of all reference symbols carrying the common reference signal within one subframe (i.e., $1$\,ms) 
and also over the measurement bandwidth of at least six physical resource blocks. 
In a typical handover measurement configuration shown in Fig.~\ref{Fig:HO_measurement_model}\subref{Fig:HO_measurement_model_David}, 
the UE performs the layer-1 filtering by obtaining an RSRP sample every $40$\,ms, 
and performs linear averaging over a number of successive RSRP samples, 
usually 5 samples.  
As a result, L1 filtering performs averaging over every $200$ ms to obtain an L1 sample, $M(n)$, 
given by~\cite{David_mag_2012}
\begin{align}
M(n) = \frac{1}{5}\sum_{k=0}^{4} RSRP_{\rm L1}(5n-k)~,
\end{align}
where $n$ is the discrete time index of the RSRP sample, $RSRP_{\rm L1}$ is the RSRP sample measured at every $40$~ms by the UE, and $k$ is the delay index of the filter.  

A UE further averages the L1 samples through first-order infinite impulse response filter 
called L3 filtering, 
which is given by
\begin{align}
F(n)=(1-a)F(n-1)+a10\log_{10}[M(n)]~,
\label{Eq:L3_filter}
\end{align}
where $a$ is the L3 filter coefficient. 

Then, UE periodically checks the resulting L3 sample every $T_{\rm d}$ seconds (e.g., $200$~ms in 3GPP LTE~\cite{Aziz_VTC_2009}) for the handover entry conditions. 
$T_{\rm d}$ refers to L3 sampling period in the 3GPP LTE standard. 
If the handover entry condition is satisfied, then rest of the handover steps may follow as described previously.


\section{A Geometric Model for the Analysis of Handover Failures}\label{Sec: analytical_model}

\begin{figure}[h]
\begin{center}
\includegraphics[width=3.5in]{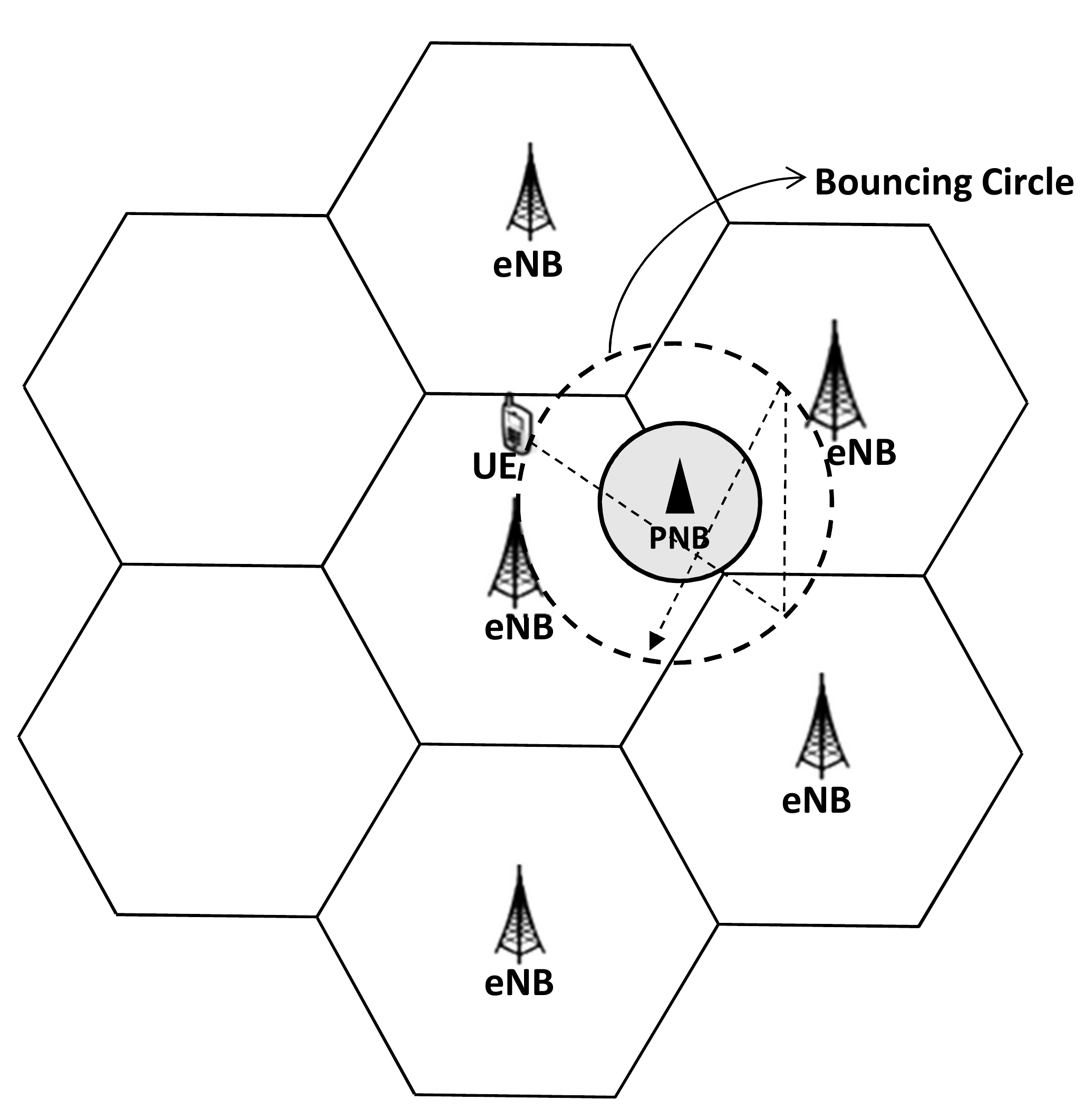}
\captionof{figure}{The bouncing ring UE mobility model from~\cite{TR36839}.}
\label{Fig:Layout_for_Hist}
\end{center}
\end{figure}

In order to evaluate the handover performance in HetNets, 
there is a standard hotspot model with specific simulation scenarios and parameters presented in the 3GPP study item~\cite{TR36839}. 
This hotspot model is based on a bouncing (hotspot) circle concentric within the picocell,
whose radius is assumed to be $200$~m. 
The starting UE position is chosen randomly on the bouncing circle and the UE follows a linear trajectory. 
The UE does not change the direction until it hits the bouncing circle and bounces back with a random angle, 
as shown in Fig.~\ref{Fig:Layout_for_Hist}. 
With such a model, theoretical analysis of the handover failure probabilities is challenging, 
due to the complexity of modeling the statistics of a UE's sojourn time with in a picocell. 
Instead, we propose a simple geometric model in the following subsection for the handover performance analysis.

\vspace{-3mm}
\subsection{New Geometric Model for Handover Performance Analysis}

In order to simplify the handover model shown in Fig.~\ref{Fig:Layout_for_Hist}, we initially consider the handover metrics of a single user in the absence of fading, and develop a new framework to facilitate closed form analysis of HF probabilities. Later, using this framework as a reference, we extend our analysis of HF probabilities into the scenario where there is channel fading.

\begin{figure}[htp]
\begin{center}
\includegraphics[width=3.5in]{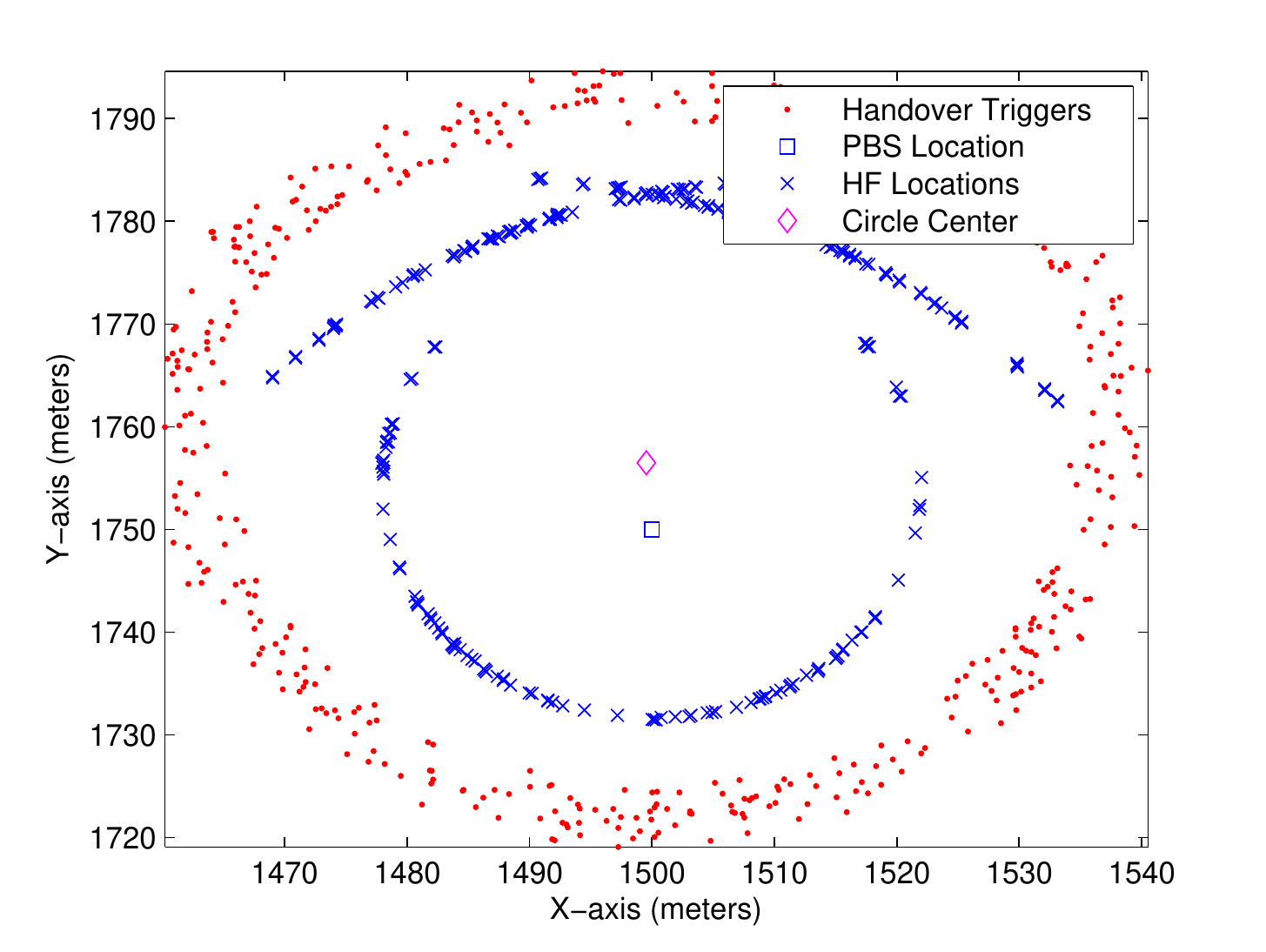}
\captionof{figure}{Handover trigger  and handover failure locations in an example picocell using a 3GPP-compliant simulator. The MBS is located at $(1500,1500)$~m and has three sectors ($T_{\rm d}=200$~ms)~\cite{David_ICC_2012}.}
\label{Fig:HO_Trigger_Locs}
\end{center}
\end{figure}

Considering the handover measurement process summarized in Fig.~\ref{Fig:HO_measurement_model}, 
and assuming that there is no fading, 
we investigate where the TTT are initiated within the coverage area of a picocell and also the handover failure locations experience by the UE. 
Using a 3GPP compliant simulator and based on the assumptions shown in Fig.~\ref{Fig:Layout_for_Hist}~\cite{David_mag_2012}, 
handover trigger (red dots, after step-1 in Fig.~\ref{Fig:HF_HetNets}) and HF locations (blue cross) are obtained. 
Fig.~\ref{Fig:HO_Trigger_Locs} shows the ring of handover trigger locations as a result of discrete measurement carried out at UE. 
These handover trigger locations extend inwards from the ideal coverage area of a pico BS (PBS), 
since a UE may delay initiation of the TTT due to the filtered RSRP measurements being available only with $200$~ms intervals. 
Fig.~\ref{Fig:HO_Trigger_Locs} also shows that if we neglect the impact of sectorized cell structure, 
the HF locations (the blue cross signs where the wideband signal to interference plus noise ratio (SINR) becomes lower than a threshold) can be approximated by a circle. 

\begin{figure}[htp]
\begin{center}
\includegraphics[width=3.5in]{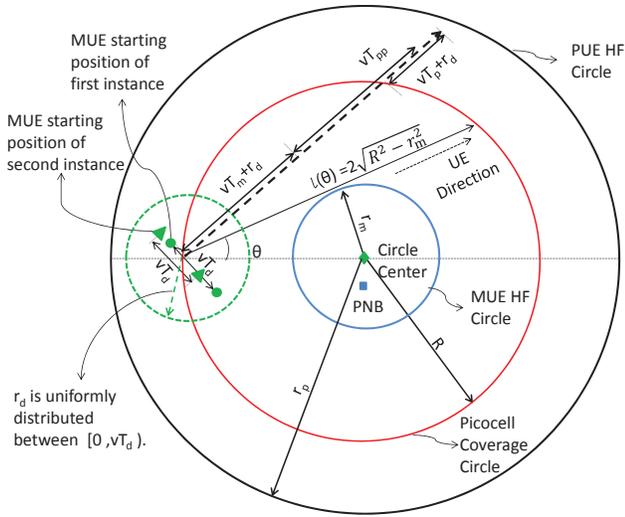}
\captionof{figure}{Geometric handover model to analyze HFs for MUEs and PUEs, and considers the effects of L3 filtering.}
\label{Fig:GeometricModel}
\end{center}
\end{figure}

Based on Fig.~\ref{Fig:HO_Trigger_Locs}, we model the picocell coverage and HF locations geometrically as concentric circles shown in Fig.~\ref{Fig:GeometricModel}, with radius $R$ and $r_{\rm m}$ respectively.
The HF locations shown in Fig.~\ref{Fig:HO_Trigger_Locs} are for  macro to pico HFs 
and in the same way it is reasonable to approximate pico to macro HF locations as another concentric circle with $r_{\rm p}$ shown in Fig.~\ref{Fig:GeometricModel}. In Fig.~\ref{Fig:GeometricModel}, $\upsilon$ denotes UE velocity, $\theta$ is the angle of UE trajectory with respect to horizontal axis, $T_{\rm m}$, $T_{\rm p}$ are the TTT duration for MUE and PUE respectively. 
In the following subsection, we incorporate the discrete measurements process in our geometric model by proposing a standard distribution that fits to the handover trigger locations. 
In the subsequent subsection, 
we use it to model the UE's sojourn time in order to facilitate the theoretical analysis of macro-cell UE (MUE) and pico-cell UE (PUE) HF.

\subsection{Modeling the Handover Trigger Locations}\label{Subsec: HO_trigger_model}
To model the statistics of the handover trigger locations which are offset due to L1/L3 filtering, we examine the distance of each handover trigger location shown in Fig.~\ref{Fig:HO_Trigger_Locs} from the ideal picocell coverage boundary,
which we define as handover offset distance. In this paper, we model the distribution of the handover offset distance using a uniform distribution for the scenario with no fading, and using a custom distribution for the scenario with fading (to be studied in Section~\ref{Sec: HF_analysis_fading}). 

We incorporate the discrete measurement process in our geometric model through modeling the handover trigger locations which is explained Fig. \ref{Fig:GeometricModel}. Let us consider a UE traveling with a velocity $\upsilon$. It is assumed that the UE is traveling in a straight line and due to the discrete measurements performed by the UE every $T_{\rm d}$ seconds, it checks whether the handover condition is satisfied at integer multiples of $T_{\rm d}$. Therefore, when the UE crosses the picocell coverage circle, TTT is not initiated immediately. It has to wait for the remaining duration of time $T_{\rm d}$ seconds depending on where the UE processed the measurement before crossing into the picocell coverage area. Let us consider different starting instance of UE crossing the picocell coverage area shown in Fig. \ref{Fig:GeometricModel}. In the first instance, the UE starts as an MUE at green dot and travels through the picocell coverage circle with a distance $\upsilon T_{\rm d}$. We can see that the TTT is triggered after the distance $\upsilon T_{\rm d}$. In the second instance the MUE starts earlier than the first instance.  Therefore, after a distance $\upsilon T_{\rm d}$, the location where the TTT would be triggered after L1/L3 filtering will be closer to the ideal picocell coverage boundary than the first instance. In this paper, we model the distance between the triggering location of the TTT and the picocell coverage circle as a random variable, and we denote this distance as $r_{\rm d}$. If we consider all possible instances, 
we assume that the distance from the cell edge reference point is uniformly distributed, i.e. $r_{\rm d}\sim\mathcal{U}[0,\upsilon T_{\rm d})$, which models the handover trigger locations.

To verify our model, 
we check the histograms for handover offset distance $r_{\rm d}$ for the trigger locations generated by the 3GPP-compliant simulator. 
Based on the assumptions shown in Fig.~\ref{Fig:Layout_for_Hist} and handover parameters shown in Table \ref{Tab:HO_parameters}, 
the handover trigger locations are generated for the following three cases:
\begin{enumerate}
\item Without shadowing and fast-fading (Case-$1$),
\item With shadowing, but without fast-fading (Case-$2$),
\item With shadowing and fast-fading (Case-$3$).
\end{enumerate}

\begin{table} [h]
\captionof{table}{Handover parameter sets.}
\begin{center}
	\begin{tabular}{ |l|l|l|l|l| }
	\hline
    Profile & Set-1 & Set-2 & Set-3 & Set-4 \\ \hline
    TTT (ms) & 480 & 160 & 80 & 40 \\ \hline
    L3 filter coefficient (a) & 4 & 1 & 1 & 0 \\
    \hline
    \end{tabular}
    \label{Tab:HO_parameters}
\end{center}
\end{table}

\begin{figure}[h]
\begin{center}
\includegraphics[width=3.5in]{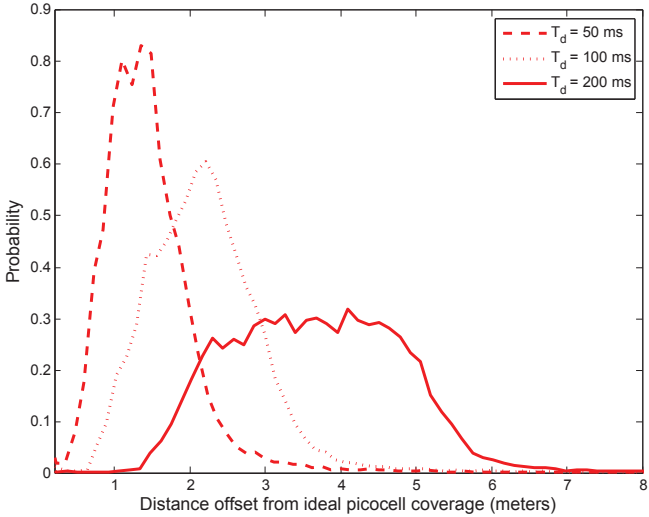}
\captionof{figure}{Handover offset distance histograms for $\upsilon = 60$~km/hr in Case-$1$. For the scenario of $T_{\rm d} = 200$~ms adopted in 3GPP LTE, the handover offset distance histogram can be reasonably modeled using a uniform distribution.}
\label{Fig:histograms_nn}
\end{center}
\end{figure}

The handover offset distance histograms for Case-$1$ is shown in Fig.~\ref{Fig:histograms_nn}, 
and we can see that for higher UE velocity the probability density function (PDF) flattens according to our proposed uniform discrete assumption, 
as described earlier. 
Modeling the handover process in the fading scenario follows from Case-$1$, and is achieved by finding the handover offset distance histograms for the trigger locations in the Case-$2$ and Case-$3$, 
which is studied in Section \ref{Sec: HF_analysis_fading}.

\subsection{Modeling the UEs' Sojourn Times}

The sojourn time estimated in a picocell may be different when using the geometric model in Fig.~\ref{Fig:GeometricModel} or when using the bouncing ring model of Fig.~\ref{Fig:Layout_for_Hist}. 
In order to justify the use of our geometric model in Fig.~\ref{Fig:GeometricModel} to model the scenario in Fig.~\ref{Fig:Layout_for_Hist}, 
we consider the Bertrand's Paradox~\cite{MIT_Website} and the three probability density functions described therein. 
We will show that one of these probability density functions well matches the behaviour of the bouncing ring model, 
thus validating our modelling.

The Bertrand's Paradox studies the probability that a random chord of a circle with a radius $R$ is larger than a threshold parameter. 
In essence, this probability leads us to the statistics of the sojourn time in a given picocell. 
Due to different interpretation of \emph{randomness} of a chord in a circle, 
there are three different models for the PDF of the chord length.

\paragraph*{\underline{Model 1}:} 
When we choose randomly two points on a circle and draw the chord joining them, 
without loss of generality, 
we may position ourselves at one of them and examine the relative location of the other points. 
If angle $\theta$ is uniformly distributed between $[-\frac{\pi}{2},\frac{\pi}{2}]$, 
then the PDF of the chord length $l$ is given by:
\begin{align}
f_{\rm 1}(l) = \frac{2}{\pi\sqrt{4R^2-l^2}}~.
\label{Eq:pdf_chord_length_case1}
\end{align}
Note that this interpretation corresponds to the model used in Fig.~\ref{Fig:GeometricModel}.

\paragraph*{\underline{Model 2}:} 
If we choose a chord whose direction is fixed and perpendicular to a given diameter of the circle, 
then we assume that the point of intersection of the chord with the diameter has a uniform distribution. 
Therefore we can assume that the perpendicular distance $r=\sqrt{R^2-\frac{d^2}{4}}$ from the chord to the center of the circle is uniformly distributed between $[0,R]$. 
The PDF of the chord length $l$ is then given by:
\begin{align}
f_{\rm 2}(l) = \frac{l}{2R\sqrt{4R^2-l^2}}~.
\label{Eq:pdf_chord_length_case2}
\end{align}

\paragraph*{\underline{Model 3}:} 
A chord is uniquely determined by its midpoint, 
for which a perpendicular line extending from the circle center intersects with the chord. 
We assume this intersection point is uniformly distributed over the entire circle and PDF of chord length is given by:
\begin{align}
f_{\rm 3}(l) = \frac{l}{2R^2}~.
\label{Eq:pdf_chord_length_case3}
\end{align}

\begin{figure}
\begin{center}
\includegraphics[width=3.5in]{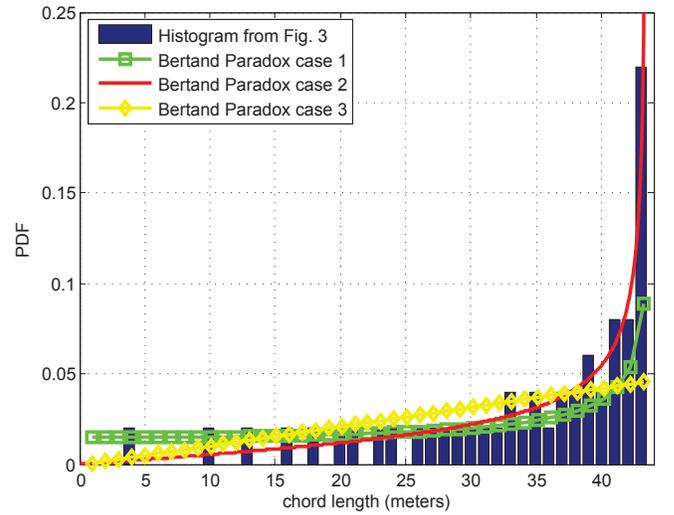}
\captionof{figure}{PDFs of the chord length for the three different approaches in Bertrand's Paradox, and the histogram of the chord length for the bouncing-ring simulations in Fig. \ref{Fig:Layout_for_Hist}.}
\label{Fig:Pdf_Bertrand_123}
\end{center}
\vspace{-8mm}
\end{figure}

In order to evaluate how closely the three approaches in the Bertrand's Paradox capture the picocell sojourn time in Fig. \ref{Fig:Layout_for_Hist}, 
we compare the PDFs of chord lengths for the three Bertrand's Paradox cases, 
with the simulated chord length histogram for the bouncing ring model in Fig.~\ref{Fig:Layout_for_Hist}. 
Considering $R=21.7$\,m and plotting the histograms of chord length overlayed with PDFs of all the three solutions in (\ref{Eq:pdf_chord_length_case1})-(\ref{Eq:pdf_chord_length_case3}), 
we obtain the results in Fig.~\ref{Fig:Pdf_Bertrand_123}. 
Model 1 shows a reasonable match with the histograms and therefore we use the PDF given in (\ref{Eq:pdf_chord_length_case1}) for our MUE and PUE HF analysis in both no fading and fading scenarios.

\section{Handover Failure Analysis without Fading}\label{Sec: HF_analysis}

Using the geometric model in Fig.~\ref{Fig:GeometricModel} and the PDF of chord length in~\eqref{Eq:pdf_chord_length_case1}, 
in this section, 
we derive the handover failure probabilities for MUEs and PUEs considering L3 filtering for the ideal handover case. 
We consider that a UE checks the handover entry condition at every $T_{\rm d}$ sampling period of the L3 filter. 
When the handover condition is satisfied, 
then TTT of duration $T_{\rm m}$ is triggered. 
The $r_{\rm d}$ is a random variable accounting for discrete measurement interval carried out in the UE through L3 filtering, 
and we assume that $r_{\rm d}$ is uniformly distributed between $[0,\upsilon T_{\rm d}]$, 
yielding the following PDF
\begin{align}
f(r_{\rm d}) =
\begin{cases}
\frac{1}{\upsilon T_{\rm d}} & 0\leq r_{\rm d}<\upsilon T_{\rm d}\\
0 & \rm otherwise\\
\end{cases}~.\label{Eq:pdf_r_d}
\end{align}

In other words, $r_{\rm d}$ models the random offset between the intersection of the UE trajectory with the ideal picocell coverage circle, and the location when the filtered measurements become available to the UE after entering the picocell's coverage area. In the following section, 
we derive the no handover (NHO) probability for MUEs and HF probabilities for both MUEs and PUEs 
using Fig.~\ref{Fig:GeometricModel}, and equations~(\ref{Eq:pdf_chord_length_case1}) and~(\ref{Eq:pdf_r_d}).

\subsection{No Handover Probability for MUEs}\label{Sec: NHO_probability}

After TTT of duration $T_{\rm m}$ is triggered, 
the MUE does not make a handover if it leaves the picocell coverage circle before the end of $T_{\rm m}$. 
In Fig.~\ref{Fig:GeometricModel} we can see that the MUE does not make a handover if $\upsilon T_{\rm m} + r_{\rm d}$ is larger than the chord length $l(\theta)=2R\cos(\theta)$, 
and for which the chord does not intersect with the MUE HF circle. 
Depending on the value of $\upsilon T_{\rm m}$ relative to $r_{\rm d}$ and picocell coverage, 
NHO probability for a given UE can be analyzed for three different cases.
\paragraph*{1)} $\upsilon T_{\rm m}< 2\sqrt{R^2-r_{\rm m}^2}-\upsilon T_{\rm d}$: 
Since random variable $r_{\rm d}$ and $\theta$ are independent, 
we can just multiply the two PDFs in (\ref{Eq:pdf_chord_length_case1}) and (\ref{Eq:pdf_r_d}) to obtain the joint PDF. 
Then, after some manipulation, 
NHO probability is expressed as
\vspace{-1mm}
\begin{align}
&\mathcal{P}_{\rm NHO}=\mathbb{P}\Big(l(\theta)<\upsilon T_{\rm m}+r_{\rm d}\Big) \nonumber \\
&= \smashoperator{\int_{0}^{\upsilon T_{\rm m}}}\!\!\frac{2}{\pi\sqrt{4R^2-l^2}} {\rm d}l\nonumber 
+\smashoperator{\int_{0}^{\upsilon T_{\rm d}}}\!\!\frac{1}{\upsilon T_{\rm d}}\!\!\smashoperator{\int_{\upsilon T_{\rm m}}^{\upsilon T_{\rm m}+r_{\rm d}}}\!\!\frac{2}{\pi\sqrt{4R^2-l^2}}\,{\rm d}r\,{\rm d}r_{\rm d}~,\nonumber \\
&= \frac{2}{\pi\upsilon T_{\rm d}}\bigg[\sqrt{4R^2-(\upsilon T_{\rm m}+\upsilon T_{\rm d})^2}+(\upsilon T_{\rm m}+\upsilon T_{\rm d})\tilde{T}_{\rm d}\nonumber \\
&-\sqrt{4R^2-(\upsilon T_{\rm m})^2} -\upsilon T_{\rm m}\tan^{-1}\bigg(\frac{\upsilon T_{\rm m}}{\sqrt{4R^2-(\upsilon T_{\rm m})^2}}\bigg)\bigg]~,
\end{align}\label{Eq:NHO_Soln1}
where 
\begin{align}
\tilde{T}_{\rm d}=\tan^{-1}\bigg(\frac{\upsilon T_{\rm m}+\upsilon T_{\rm d}}{\sqrt{4R^2-(\upsilon T_{\rm m}+\upsilon T_{\rm d})^2}}\bigg)~.
\end{align}
\paragraph*{2)} $\upsilon T_{\rm m}\geq 2\sqrt{R^2-r_{\rm m}^2}-\upsilon T_{\rm d}$:
If $\upsilon T_{\rm m}+\upsilon T_{\rm d}$ is greater than the chord length $2\sqrt{R^2-r_{\rm m}^2}$, 
then we have to subtract the integration region which overlaps with the MUE HF circle from case 1, 
which yields
\begin{align}
\mathcal{P}_{\rm NHO}&=\mathbb{P}\Big(l(\theta)<\upsilon T_{\rm m}+r_{\rm d}\mid r_{\rm d}\leq 2\sqrt{R^2-r_{\rm m}^2}-\upsilon T_{\rm m}\Big)\nonumber \\
&= \!\!\smashoperator{\int_{0}^{\upsilon T_{\rm m}}}\!\!\frac{2}{\pi\sqrt{4R^2-l^2}} {\rm d}l+\smashoperator{\int_{0}^{\upsilon T_{\rm d}}}\!\!\frac{1}{\upsilon T_{\rm d}} \smashoperator{\int_{\upsilon T_{\rm m}}^{\upsilon T_{\rm m}+r_{\rm d}}}\!\!\frac{2}{\pi\sqrt{4R^2-l^2}}\,{\rm d}r\,{\rm d}r_{\rm d}\nonumber \\
&-\smashoperator{\int_{2\sqrt{R^2-r_{\rm m}^2}-\upsilon T_{\rm m}}^{\upsilon T_{\rm d}}}\!\!\frac{1}{\upsilon T_{\rm d}}\;\;\;\;\smashoperator{\int_{\upsilon T_{\rm m} + r_{\rm d}}^{\upsilon T_{\rm m}+\upsilon T_{\rm d}}}\!\!\frac{2}{\pi\sqrt{4R^2-l^2}}\,{\rm d}r\,{\rm d}r_{\rm d}~.\label{Eq:NHO_Int2}
\end{align}
The limits of the random variable $r_{\rm d}$ in the third term represent the integration region which overlaps with MUE HF circle. 
After solving the integral in (\ref{Eq:NHO_Int2}) and using $\tilde{T}_{\rm d}$ we get
\begin{align}
&\mathcal{P}_{\rm NHO}=\frac{2}{\pi\upsilon T_{\rm d}}\bigg[(\upsilon T_{\rm m}+\upsilon T_{\rm d}+2\sqrt{R^2-r_{\rm m}^2})\tilde{T}_{\rm d}-2r_{\rm m}\nonumber \\
&-\upsilon T_{\rm m}\tan^{-1}\bigg(\frac{\upsilon T_{\rm m}}{\sqrt{4R^2-(\upsilon T_{\rm m})^2}}\bigg)
+2\sqrt{4R^2-(\upsilon T_{\rm m}+\upsilon T_{\rm d})^2}\nonumber \\
&-\sqrt{4R^{2}-(\upsilon T_{\rm m})^{2}}-(2\sqrt{R^2-r_{\rm m}^2})\tan^{-1}\bigg(\frac{\sqrt{R^2-r_{\rm m}^2}}{r_{\rm m}}\bigg)\bigg]~.
\end{align}\label{Eq:NHO_Soln2}
\vspace{-2mm}
\paragraph*{3)} 
If UE velocity is too high and product $\upsilon T_{\rm m}$ is itself very large and greater than the chord length $2\sqrt{R^2-r_{\rm m}^2}$, 
then there is no effect of L3 sampling period on NHO probability. 
In this case, we obtain a constant NHO probability due to the fixed limits in $\theta$, 
and using (\ref{Eq:pdf_chord_length_case1}) the NHO probability becomes
\vspace{-5mm}
\begin{align}
\mathcal{P}_{\rm NHO}&=\mathbb{P}\Big(l(\theta)>2\sqrt{R^2-r_{\rm m}^2}\Big) =\smashoperator{\int_{0}^{2\sqrt{R^2-r_{\rm m}^2}}}\frac{2}{\pi\sqrt{4R^2-l^2}} {\rm d}l\nonumber \\
&=\frac{2}{\pi}\tan^{-1}\bigg(\frac{\sqrt{R^2-r_{\rm m}^2}}{r_{\rm m}}\bigg)~.
\end{align}\label{Eq:NHO_Soln3}
\vspace{-7mm}

\subsection{HF Probability for MUEs}

When MUE reaches the MUE HF circle before TTT expires, 
there will be handover failure and the MUE fails to send a measurement report. 
Therefore, the MUE fails to connect with the pico cell. 
From Fig. \ref{Fig:GeometricModel}, we can see that if $l(\theta)\geq 2\sqrt{R^2-r_{\rm m}^2}$, 
then the MUE trajectory intersects with the MUE HF circle and therefore there may be a MUE handover failure.

A MUE HF occurs only when the total distance traveled by MUE, 
which is $\upsilon T_{\rm m}+r_{\rm d}$, 
is greater than the distance $d_{\rm HF,m}(\theta,R,r_{\rm m})$. 
This distance refers to total distance travelled by the UE from the ideal picocell coverage to the MUE HF circle and is given by
\begin{align}
d_{\rm HF,m}(\theta,R,r_{\rm m})=R\cos(\theta)-\sqrt{r_{\rm m}^2-R^2\sin^2(\theta)}~.\label{Eq:d_hf_m}
\end{align}
To obtain the MUE HF probability, 
first we evaluate the MUE HF condition $\upsilon T_{\rm m}+r_{\rm d}>d_{\rm HF,m}(\theta,R,r_{\rm m})$ in terms of UE trajectory $l(\theta)$. 
Using (\ref{Eq:d_hf_m}), we can write
\begin{align}
\upsilon T_{\rm m}+r_{\rm d}>R\cos(\alpha)-\sqrt{r_{\rm m}^2-R^2\sin^2(\alpha)}~.
\label{Eq:MUEHF_cond_1}
\end{align}
Using $l(\theta)=2R\cos(\theta)$, 
we get $\theta=\cos^{-1}(\frac{l(\theta)}{2R})$, 
and using this in (\ref{Eq:MUEHF_cond_1}), 
we get the MUE HF condition in terms of $l(\theta)$ as
\begin{align}
l(\theta)>\frac{R^2-r_{\rm m}^2}{\upsilon T_{\rm m}+r_{\rm d}}+(\upsilon T_{\rm m}+r_{\rm d})~.
\label{Eq:MUEHF_cond_chord_length}
\end{align}
Then the MUE HF probability is calculated differently for the following four cases.

\paragraph*{1)} $\upsilon T_{\rm m}<R-r_{\rm m}-\upsilon T_{\rm d}$:
We know that for $\theta=0$, $d_{\rm HF,min}=R-r_{\rm m}$. 
If the velocity of the UE is very slow then the distance traveled by UE including the discrete L3 filter measurement offset is $\upsilon T_{\rm m}+r_{\rm d}$. 
If this distance is less than $d_{\rm HF,min}$, 
then there will be no MUE HF, i.e., $\mathcal{P}_{\rm HF,m}=0$.

\paragraph*{2)} $R-r_{\rm m}<\upsilon T_{\rm m}+\upsilon T_{\rm d}<\sqrt{R^2-r_{\rm m}^2}$:
When the velocity of the UE is high, 
then we have obtained the MUE HF condition in (\ref{Eq:MUEHF_cond_chord_length}) and we obtain the MUE HF probability as
\begin{align}
&P_{\rm HF,m} = \mathbb{P}\Big(l(\theta)>\frac{R^2-r_{\rm m}^2}{\upsilon T_{\rm m}+r_{\rm d}}+(\upsilon T_{\rm m}+r_{\rm d})\Big) \nonumber\\ 
&=\smashoperator{\int_{0}^{\upsilon T_{\rm d}}}\frac{1}{\upsilon T_{\rm d}}\!\! \int_{\frac{R^2-r_{\rm m}^2}{\upsilon T_{\rm m}+r_{\rm d}}+\upsilon T_{\rm m}+r_{\rm d}}^{\frac{R^2-r_{\rm m}^2}{\upsilon T_{\rm m}}+\upsilon T_{\rm m}}\frac{2}{\pi\sqrt{4R^2-l^2}}{\rm d}l\,{\rm d}r_{\rm d}
\nonumber\\
&+ \smashoperator{\int_{\frac{R^2-r_{\rm m}^2}{\upsilon T_{\rm m}}+\upsilon T_{\rm m}}^{2R}}\frac{2}{\pi\sqrt{4R^2-l^2}}{\rm d}l\,{\rm d}r_{\rm d} \nonumber\\ 
&=1-\int_{0}^{\upsilon T_{\rm d}}\!\!\frac{2}{\pi\upsilon T_{\rm d}} I_1(r_{\rm d}){\rm d}r_{\rm d}~,
\end{align}
where $I_1(r_{\rm d})$ is
\begin{align}
\tan^{-1}\bigg(\frac{R^2-r_{\rm m}^2+(\upsilon T_{\rm m}+r_{\rm d})^2}{\sqrt{4R^2(\upsilon T_{\rm m}+r_{\rm d})^2-(R^2-r_{\rm m}^2+(\upsilon T_{\rm m}+r_{\rm d})^2)^2}}\bigg)~.
\end{align}
\paragraph*{3)} $\upsilon T_{\rm m}>\sqrt{R^2-r_{\rm m}^2}-\upsilon T_{\rm d}$:
For this case, 
the MUE HF probability is the same as for case $2$.

\paragraph*{4)} 
If the UE velocity is too high and $\upsilon T_{\rm m}$ is very large and greater than the chord length $\sqrt{R^2-r_{\rm m}^2}$, 
then MUE HF probability is independent of the random variable $r_{\rm d}$, 
and we obtain a constant probability, 
which is the same as for case $3$ of NHO probability.

\subsection{HF Probability for PUEs}

In order to observe a PUE HF there should be a successful handover of MUE to the pico-cell. 
After a successful handover, 
the PUE starts recording the measurements of PNB for every L3 sampling period $T_{\rm d}$. 
If a PUE enters the coverage of the macrocell and handover event entry condition is satsified 
(e.g., L3 filtered RSRP of the macro cell is larger than that of the picocell plus an hysteresis), 
then TTT of duration $T_{\rm p}$ (see Fig.~\ref{Fig:GeometricModel}) is triggered. 
For simplicity, we assume that discrete offset random variable $r_{\rm d}$ is the same whenever there is a handover to picocell or macrocell.

If a PUE reaches 
the PUE HF circle before the TTT expires, 
there will be a PUE HF. 
In other words, a PUE HF occurs when the total distance travelled by PUE $(\upsilon T_{\rm p}+r_{\rm d})$ is greater than the distance $d_{\rm HF,p}(\theta,R,r_{\rm p})$. 
If we consider a point on the ideal picocell coverage area where the UE starts entering the coverage of the macrocell after a successful handover to picocell, 
then distance from this point to the PUE HF circle is given by
\begin{align}
d_{\rm HF,p}(\theta,R,r_{\rm p})=R\cos(\theta)+\sqrt{r_{\rm p}^2-R^2\sin^2(\theta)}-d(\theta)~.\label{Eq:d_hf_p}
\vspace{-5mm}
\end{align}

To obtain the PUE HF probability, 
we will evaluate the PUE HF condition $\upsilon T_{\rm p}+r_{\rm d}>d_{\rm HF,p}(\theta,R,r_{\rm p})$ in terms of UE trajectory $l(\theta)$ 
like we did before for MUE HF. 
Using (\ref{Eq:d_hf_p}), 
we get the condition for observing PUE HF as
\begin{align}
l(\theta)>\frac{r_{\rm p}^2-R^2}{\upsilon T_{\rm p}+r_{\rm d}}-(\upsilon T_{\rm p}+r_{\rm d})~.
\label{Eq:PUEHF_cond_chord_length}
\end{align}

Based on the condition in (\ref{Eq:PUEHF_cond_chord_length}) and the condition $l(\theta)>2\sqrt{R^2-r_{\rm m}^2}$, 
we can show that when $\upsilon T_{\rm m}+\upsilon T_{\rm p}>\sqrt{r_{\rm p}^2-r_{\rm m}^2}-\sqrt{R^2-r_{\rm m}^2}$, 
there will be a PUE HF. For different values of $\upsilon T_{\rm m}$ and $\upsilon T_{\rm d}$, 
the PUE HF is given as follows:

\paragraph*{1)} $\upsilon T_{\rm m}+\upsilon T_{\rm d}<\sqrt{R^2-r_{\rm m}^2}$:
In this case, in order to observe PUE HF, 
we have to make sure that there is no MUE handover failure and we will use MUE HF condition in (\ref{Eq:MUEHF_cond_chord_length}) and PUE HF condition in (\ref{Eq:PUEHF_cond_chord_length}) to obtain PUE HF. 
We can see that there is no MUE HF even when $l(\theta)>2\sqrt{R^2-r_{\rm m}^2}$.
 Moreover,  there will be a possible PUE HF for the condition 
\begin{align} 
 \upsilon T_{\rm m}+\upsilon T_{\rm p}>\sqrt{r_{\rm p}^2-r_{\rm m}^2}-\sqrt{R^2-r_{\rm m}^2}~.
 \end{align} 
After some manipulation PUE HF probability is given by
\begin{align}
P_{\rm HF,p}&=\mathbb{P}\Bigg(\max\bigg(\upsilon T_{\rm m}+r_{\rm d},\frac{r_{\rm p}^2-R^2}{\upsilon T_{\rm p}+r_{\rm d}}-(\upsilon T_{\rm p}+r_{\rm d})\bigg)\nonumber \\
&<d(\alpha)< \min\bigg(\frac{R^2-r_{\rm m}^2}{\upsilon T_{\rm m}+r_{\rm d}}+(\upsilon T_{\rm m}+r_{\rm d}),2R\bigg)\Bigg)~.
\end{align}
Consider the following definitions for brevity: 
\begin{align}
&d_{\rm rp} = \frac{r_{\rm p}^2-R^2}{\upsilon T_{\rm p}+r_{\rm d}}-(\upsilon T_{\rm p}+r_{\rm d})~,\label{Eq: drp}\\
&d_{\rm rm} = \frac{R^2-r_{\rm m}^2}{\upsilon T_{\rm m}+r_{\rm d}}+(\upsilon T_{\rm m}+r_{\rm d})~,\label{Eq: drm}\\
&d_{\rm vm}=\upsilon T_{\rm m}+r_{\rm d}~.\label{Eq: dvtm}
\end{align}
Using (\ref{Eq: drp})--(\ref{Eq: dvtm}) we can then write $L_{\rm p} = \max(d_{\rm vm}, d_{\rm rp})$ and $L_{\rm m} = \min(d_{\rm rm},2R)$. Using the PDF in (\ref{Eq:pdf_chord_length_case1}), we can calculate the PUE HF probability after some manipulation as
\begin{align}
&P_{\rm HF,p} = \int_{0}^{\upsilon T_{\rm d}}\frac{1}{\upsilon T_{\rm d}}\!\! \int_{L_{\rm p}}^{L_{\rm m}} \frac{2}{\pi\sqrt{4R^2-l^2}}{\rm d}l\,{\rm d}r_{\rm d} \nonumber \\
&= \smashoperator{\int_{0}^{R-r_{\rm m}-\upsilon T_{\rm m}}}\frac{1}{\upsilon T_{\rm d}}{\rm d}r_{\rm d} + \smashoperator{\int_{R-r_{\rm m}-\upsilon T_{\rm m}}^{\upsilon T_{\rm d}}}\frac{2}{\pi\upsilon T_{\rm d}} \tan^{-1}\bigg(\frac{d_{\rm rm}}{\sqrt{4R^2-d_{\rm rm}^2}}\bigg){\rm d}r_{\rm d} 
\nonumber \\
&- \smashoperator{\int_{0}^{l_{\rm p}}} \tan^{-1}\bigg(\frac{d_{\rm rp}}{\sqrt{4R^2-d_{\rm rp}^2}}\bigg){\rm d}r_{\rm d} - \smashoperator{\int_{l_{\rm p}}^{\upsilon T_{\rm d}}} \tan^{-1}\bigg(\frac{d_{\rm vm}}{\sqrt{4R^2-d_{\rm vm}^2}}\bigg){\rm d}r_{\rm d}~,
\end{align}
where $l_{\rm p}$ is
\begin{align}
\frac{\sqrt{-8R^2+8r_{\rm p}^2+(\upsilon T_{\rm m})^2-2\upsilon^{2}T_{\rm m}T_{\rm p}+(\upsilon T_{\rm p})^2-3\upsilon^{2}T_{\rm m}T_{\rm p}}}{4}~.
\end{align}

\paragraph*{2)} $\sqrt{R^2-r_{\rm m}^2}<\upsilon T_{\rm m}+\upsilon T_{\rm d}<2\sqrt{R^2-r_{\rm m}^2}$:
In this case, MUE HF occurs when $l(\theta)>2\sqrt{R^2-r_{\rm m}^2}$.
Using this condition and the PUE HF condition in (\ref{Eq:PUEHF_cond_chord_length}), 
the PUE HF probability is given by
\vspace{-2mm}
\begin{align}
P_{\rm HF,p}=\mathbb{P}\Bigg(\max\bigg(\upsilon T_{\rm m}+r_{\rm d},\frac{r_{\rm p}^2-R^2}{\upsilon T_{\rm p}+r_{\rm d}}-(\upsilon T_{\rm p}+r_{\rm d})\bigg)\nonumber \\
<d(\alpha)<2\sqrt{R^2-r_{\rm m}^2}\Bigg)~.\label{Eq: PUEHF_prob_2}
\end{align}
Using (\ref{Eq:pdf_chord_length_case1}), (\ref{Eq: drp})--(\ref{Eq: dvtm}), $L_{\rm m}$ and $L_{\rm p}$,
we can find PUE HF probability after some manipulation as
\begin{align}
&P_{\rm HF,p} = \int_{0}^{\upsilon T_{\rm d}}\frac{1}{\upsilon T_{\rm d}}\!\! \int_{L_{\rm p}}^{r_{\rm m}} \frac{2}{\pi\sqrt{4R^2-l^2}}{\rm d}l\,{\rm d}r_{\rm d} \nonumber \\
&= \frac{2}{\pi}\tan^{-1}\bigg(\frac{\sqrt{R^2-r_{\rm m}^2}}{r_{\rm m}}\bigg)+ \smashoperator{\int_{0}^{l_{\rm p}}} \tan^{-1}\bigg(\frac{d_{\rm rp}}{\sqrt{4R^2-d_{\rm rp}^2}}\bigg){\rm d}r_{\rm d} \nonumber \\
&+ \smashoperator{\int_{l_{\rm p}}^{\upsilon T_{\rm d}}} \tan^{-1}\bigg(\frac{d_{\rm vm}}{\sqrt{4R^2-d_{\rm vm}^2}}\bigg){\rm d}r_{\rm d}~.
\end{align}

\paragraph*{3)} $\upsilon T_{\rm m}+\upsilon T_{\rm d}>2\sqrt{R^2-r_{\rm m}^2}$:
In this case, PUE HF is the same as in case $2$ of PUE HF probability.
\vspace{-4mm}

\section{Handover Failure Analysis with Fading}\label{Sec: HF_analysis_fading}

The geometrical model framework to perform handover failure analysis in the fading scenario is similar compared to no fading scenario. We model the \emph{handover failure} locations in the fading scenario as a circle, because the wideband SINR (which dictates handover failures) is typically averaged over a large bandwidth, and the effect of fading is minimal. On the contrary, the \emph{handover trigger} locations in the fading scenario depend on the RSRP, which is measured over a narrow bandwidth of six resource blocks in LTE. Therefore, RSRP is subject to much larger randomness, and hence, may significantly affect the handover trigger locations. In the following section, we explain the modeling of handover trigger location in the fading scenario and investigate its distribution to facilitate the theoretical analysis of handover failure probability.
\subsection{Modeling the Handover Trigger Locations in the Fading Scenario}\label{Subsec: HO_trigger_model_fading}

\begin{figure}[h]
\begin{center}
\includegraphics[width=3.25in]{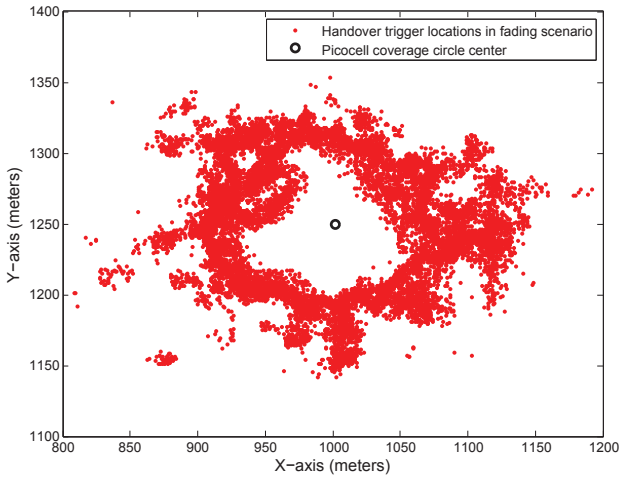}
\captionof{figure}{Handover locations in the fast-fading and shadowing scenario (Case~3) for UE velocity $\upsilon = 60$~ km/h and $T_{\rm d} = 40$~ms.}
\vspace{-5mm}
\label{Fig:HO_Trigger_Locs_YY}
\end{center}
\end{figure}

\begin{figure}[h]
\begin{center}
\includegraphics[width=3.25in]{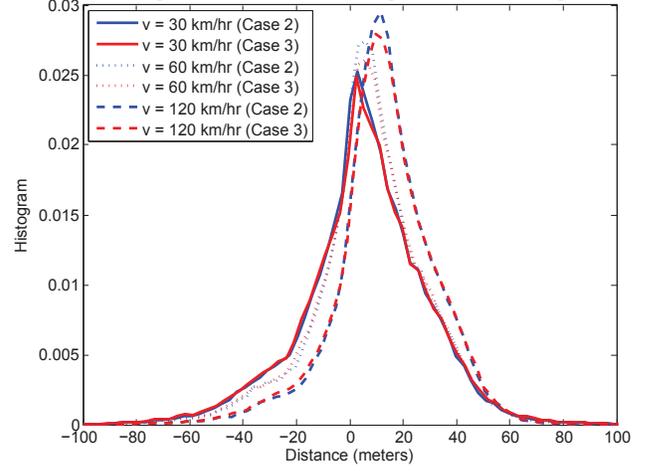}
\captionof{figure}{Handover offset distance histograms for $\upsilon = 120$~ km/h in fading scenario.}
\label{Fig:histograms_yy_yn}
\end{center}
\end{figure}

Modeling the handover trigger locations in the fading scenario is performed by examining the handover trigger locations shown in Fig.~\ref{Fig:HO_Trigger_Locs_YY}.
Comparing Fig. \ref{Fig:HO_Trigger_Locs_YY} with Fig. \ref{Fig:HO_Trigger_Locs}, we can see that due to fading channel conditions handover trigger locations in Fig. \ref{Fig:HO_Trigger_Locs_YY} are not delimited within the coverage boundaries of the picocell. In other words, TTT timer can be initiated for locations that are far away from the ideal picocell coverage area.  To model the statistics of the handover trigger locations with L1/L3 filtering, we examine the distance of each handover trigger location shown in Fig.~\ref{Fig:HO_Trigger_Locs_YY} from the ideal picocell coverage boundary which we define as handover offset distance. We model this distance using a random variable denoted by $\hat{r}_{\rm d}$ and obtain histograms for it as shown in Fig.~\ref{Fig:histograms_yy_yn}. The negative distance in histograms is due to the possibility of handover trigger locations being outside the ideal picocell coverage area due to fading.
\vspace{-5mm}
\vspace{-1mm}
After testing several standard distributions, we conclude that there is no standard distribution that reasonably approximates the histograms in  Fig.~\ref{Fig:histograms_yy_yn}. 
Therefore, for the fading scenario, we directly use the histogram data to obtain the semi-analytic handover failure expressions for MUEs and PUEs. The histograms for $\hat{r}_{\rm d}$ is given by $f(\hat{r}_{\rm d})$, where $\hat{r}_{\rm d}$ $\in(\hat{r}_{\rm min},\hat{r}_{\rm max})$.
\vspace{-1mm}

\subsection{MUE HF Probability Analysis for MUEs}

The derivation of MUE HF probabilities in shadowing and fast-fading is carried in a similar manner to that of the ideal handover model case. 
The only difference comes with distribution of the random variable $\hat{r}_{\rm d}$, 
which is obtained directly from the histograms as discussed in Section \ref{Subsec: HO_trigger_model_fading}. The MUE HF probability for a given UE can be analyzed as 
\begin{align}
&\mathbb{P}\Big(l(\theta)>\frac{R^2-r_{\rm m}^2}{\upsilon T_{\rm m}+\hat{r}_{\rm d}}+(\upsilon T_{\rm m}+\hat{r}_{\rm d})\Big) =\smashoperator{\int\limits_{\hat{r}_{\rm min}}^{\hat{r}_{\rm max}}} f(\hat{r}_{\rm d}){\rm d}\hat{r}_{\rm d} \nonumber\\ 
&\smashoperator{\int\limits_{\hat{d}_{\rm rm}}^{2R}}\frac{2}{\pi\sqrt{4R^2-l^2}}{\rm d}l = \smashoperator{\int\limits_{\hat{r}_{\rm min}}^{\hat{r}_{\rm max}}} f(\hat{r}_{\rm d})d\hat{r}_{\rm d} - \smashoperator{\int\limits_{\hat{r}_{\rm min}}^{\hat{r}_{\rm max}}} I_1(\hat{r}_{\rm d})d\hat{r}_{\rm d} \nonumber\\
&= \smashoperator{\int\limits_{\hat{r}_{\rm min}}^{\hat{r}_{\rm max}}} f(\hat{r}_{\rm d})d\hat{r}_{\rm d} - \smashoperator{\int\limits_{R-r_{\rm m}-\upsilon T_{\rm m}}^{\sqrt{R^2-r_{\rm m}^{2}}-\upsilon T_{\rm m}}} I_1(\hat{r}_{\rm d})f(\hat{r}_{\rm d})d\hat{r}_{\rm d} \nonumber\\
&-\smashoperator{\int\limits_{\sqrt{R^2-r_{\rm m}^{2}}-\upsilon T_{\rm m}}^{\hat{r}_{\rm max}}}\tan^{-1}\bigg(\frac{\sqrt{R^2-r_{\rm m}^{2}}}{r_{\rm m}}\bigg)f(\hat{r}_{\rm d})d\hat{r}_{\rm d} - \smashoperator{\int\limits_{\hat{r}_{\rm min}}^{R-r_{\rm m}-\upsilon T_{\rm m}}} f(\hat{r}_{\rm d})d\hat{r}_{\rm d}~,
\end{align}
where $I_1(\hat{r}_{\rm d})$ is
\begin{align}
\tan^{-1}\bigg(\frac{R^2-r_{\rm m}^2+(\upsilon T_{\rm m}+\hat{r}_{\rm d})^2}{\sqrt{4R^2(\upsilon T_{\rm m}+\hat{r}_{\rm d})^2-(R^2-r_{\rm m}^2+(\upsilon T_{\rm m}+\hat{r}_{\rm d})^2)^2}}\bigg)~.
\end{align} 

In Section~\ref{Sec: simulation_results}, the above integrals are solved numerically to provide numerical results for handover failure in the presence of fading.

\subsection{PUE HF Probability Analysis for PUEs}
The derivation of PUE HF probabilities in shadowing and fast-fading is carried out in a similar manner to that of the ideal handover model. The only difference is in the  distribution of the random variable $\hat{r}_{\rm d}$, 
which is obtained directly from the histograms as discussed in Section \ref{Subsec: HO_trigger_model_fading}. Using MUE HF condition in (\ref{Eq:MUEHF_cond_chord_length}), PUE HF condition in (\ref{Eq:PUEHF_cond_chord_length}), and after some manipulation, 
the PUE HF probability is expressed as
\begin{align}
P_{\rm HF,p}&=\mathbb{P}\Bigg(\max\bigg(\upsilon T_{\rm m}+\hat{r}_{\rm d},\frac{r_{\rm p}^2-R^2}{\upsilon
T_{\rm p}+r_{\rm d}}-(\upsilon T_{\rm p}+\hat{r}_{\rm d})\bigg) \nonumber \\
&<d(\alpha)<\min\bigg(\frac{R^2-r_{\rm m}^2}{\upsilon T_{\rm m}+\hat{r}_{\rm d}}+(\upsilon T_{\rm m}+\hat{r}_{\rm d}),2R\bigg)\Bigg)~.
\end{align}\label{Eq: PUEHF_prob_1}
Consider the following definitions for brevity: 
\begin{align}
 &\hat{d}_{\rm rp} = \frac{r_{\rm p}^2-R^2}{\upsilon T_{\rm p}+\hat{r}_{\rm d}}-(\upsilon T_{\rm p}+\hat{r}_{\rm d})~,\label{Eq: drp_hat}\\ 
&\hat{d}_{\rm rm} = \frac{R^2-r_{\rm m}^2}{\upsilon T_{\rm m}+\hat{r}_{\rm d}}+(\upsilon T_{\rm m}+\hat{r}_{\rm d})~,\label{Eq: drm_hat}\\
&\hat{d}_{\rm vm}=\upsilon T_{\rm m}+\hat{r}_{\rm d}~.\label{Eq: dvm_hat}
\end{align} 
Then, using (\ref{Eq: drp_hat})--(\ref{Eq: dvm_hat}) we can write $\hat{L}_{\rm p} = \max(d_{\rm vm}, d_{\rm rp})$ and $\hat{L}_{\rm m} = \min(d_{\rm rm},2R).$
 Using the PDF in (\ref{Eq:pdf_chord_length_case1}), we can calculate the PUE HF probability as 
\begin{align}
P_{\rm HF,p} &= \smashoperator{\int_{\hat{r}_{\rm min}}^{\hat{r}_{\rm max}}}f(\hat{r}_{\rm d})\!\! \smashoperator{\int_{L_{\rm p}}^{L_{\rm m}}} \frac{2}{\pi\sqrt{4R^2-l^2}}{\rm d}l\,{\rm d}\hat{r}_{\rm d} \nonumber \\
&= \smashoperator{\int_{\hat{r}_{\rm min}}^{R-r_{\rm m}-\upsilon T_{\rm m}}}\frac{1}{\upsilon T_{\rm d}}{\rm d}\hat{r}_{\rm d} -\smashoperator{\int_{2R-\upsilon T_{\rm m}}^{\hat{r}_{\rm max}}}f(\hat{r}_{\rm d}){\rm d}\hat{r}_{\rm d}- \smashoperator{\int_{\hat{r}_{\rm min}}^{r_{\rm p}-R-\upsilon T_{\rm p}}}f(\hat{r}_{\rm d}){\rm d}\hat{r}_{\rm d} \nonumber \\
&-\smashoperator{\int_{l_{\rm p}}^{2R-\upsilon T_{\rm m}}}\tan^{-1}\bigg(\frac{d_{\rm vm}}{\sqrt{4R^2-d_{\rm vm}^2}}\bigg){\rm d}\hat{r}_{\rm d} \nonumber \\
&-\smashoperator{\int_{r_{\rm p}-R-\upsilon T_{\rm p}}}^{l_{\rm p}}\tan^{-1}\bigg(\frac{\hat{d}_{\rm rp}}{\sqrt{4R^2-\hat{d}_{\rm rp}^2}}\bigg){\rm d}\hat{r}_{\rm d}\nonumber\\
&+ \smashoperator{\int\limits_{\sqrt{R^2-r_{\rm m}^{2}}-\upsilon T_{\rm m}}^{\hat{r}_{\rm max}}}\tan^{-1}\bigg(\frac{\sqrt{R^2-r_{\rm m}^{2}}}{r_{\rm m}}\bigg)f(\hat{r}_{\rm d}){\rm d}\hat{r}_{\rm d}\nonumber\\
&+\smashoperator{\int_{R-r_{\rm m}-\upsilon T_{\rm m}}^{\sqrt{R^2-r_{\rm m}^{2}}-\upsilon T_{\rm m}}}\frac{2}{\pi\upsilon T_{\rm d}} \tan^{-1}\bigg(\frac{\hat{d}_{\rm rm}}{\sqrt{4R^2-\hat{d}_{\rm rm}^2}}\bigg){\rm d}\hat{r}_{\rm d}~.
\end{align}
Again, to obtain numerical results in Section~\ref{Sec: simulation_results}, the above integrals are solved numerically using the histograms $f(r_d)$ from Fig.~\ref{Fig:histograms_yy_yn}.
%

\section{Numerical Results}\label{Sec: simulation_results}

In order to validate our analysis, 
computer simulations are carried out in Matlab. 
Initially, as shown in Fig~.\ref{Fig:GeometricModel}, we fix the starting position of MUE to a reference point on the picocell coverage circle. 
MUE travels a distance equal to $\upsilon T_{\rm m}+r_{\rm d}$ from the reference point with different realizations of angle $\theta$, 
which is uniformly distributed in $[-\frac{\pi}{2},\frac{\pi}{2}]$. 
Since $r_{\rm d}$ is uniformly distributed between $[0,\upsilon T_{\rm d}]$, 
we have different values of $r_{\rm d}$ for each $\theta$. 
After MUE travels a distance equal to $\upsilon T_{\rm m}+r_{\rm d}$, 
the final point of the MUE is checked for intersection with MUE HF circle. 
If it intersects with the MUE HF circle then there is a MUE HF. 
We aggregate all these MUE HF events and normalize over $\theta$ and $r_{\rm d}$ to obtain MUE HF probabilities for each UE velocity.

In order to obtain PUE HF probability, 
we find end points of the chord using $l(\theta)=2R\cos(\theta)$ for the picocell coverage circle for different $\theta$ realizations. 
Then, we take those points as the reference point and fix the starting position for the PUE. Next the PUE travels a distance equal to $\upsilon T_{\rm p}+r_{\rm d}$ and the final point of the PUE is checked for intersection with PUE HF circle. 
If it intersects with the PUE HF circle then there is a PUE HF. 
We aggregate all these PUE HF events and divide with the number of $\theta$ and $r_{\rm d}$ realizations to calculate PUE HF probabilities. In the following subsections the MUE and PUE HF probabilities are shown for $R=64$\,m, $r_{\rm m}=50$\,m and $r_{\rm p}=78$\,m in the no fading and fading scenarios.

\subsection{Results with No Fading}

\begin{figure}[h]
\begin{center}
\subfloat[MUE HF probability with no fading.]{\includegraphics[width=3.25in]{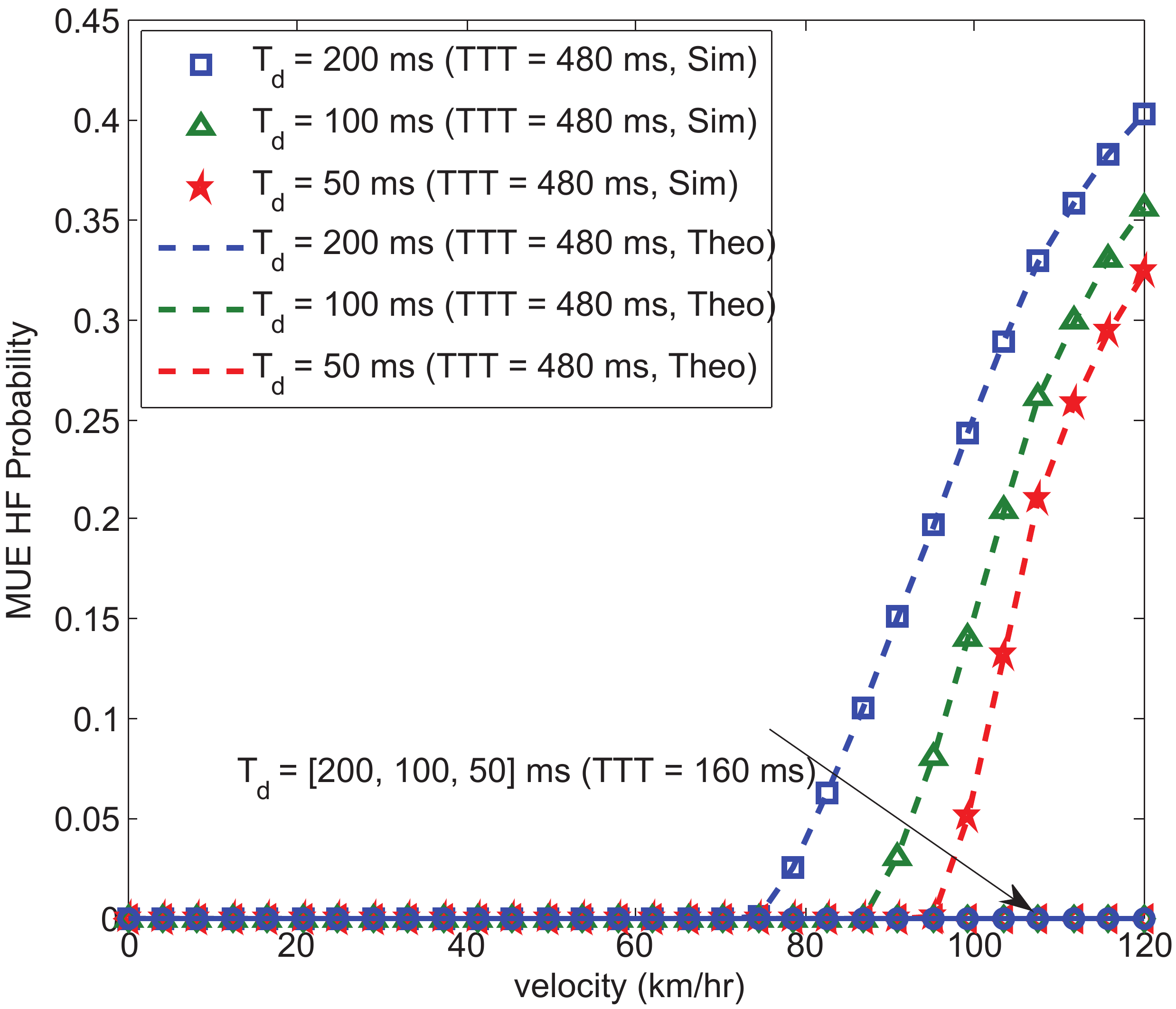}
\label{Fig:MUEHF probability_nofad}}\\
\subfloat[PUE HF probability with no fading.]{\includegraphics[width=3.25in]{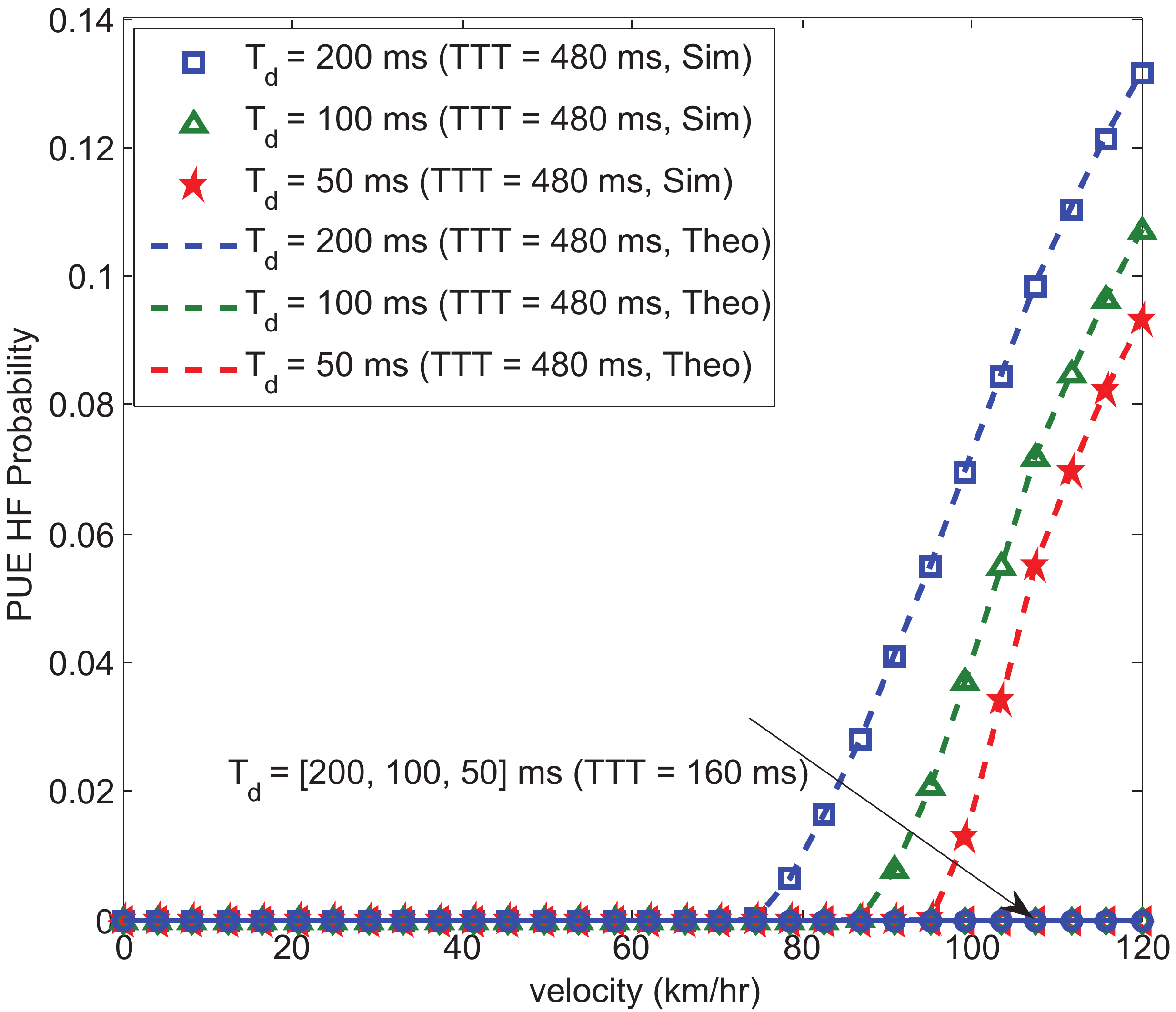}
\label{Fig:PUEHF probability_nofad}}
\captionof{figure}{Theoretical (lines) and simulated (markers) 	                                  no fading results as a function of UE velocity for $R=64$~m, $r_{\rm m}=50$~m and $r_{\rm p}=78$~m.}
\label{Fig:No_fading_res}
\end{center}
\end{figure}

Using the above simulation assumptions, 
theoretical MUE HF and PUE HF probabilities derived in Section~\ref{Sec: HF_analysis} are plotted as a function of UE velocity and are verified via simulation results. Initially, it is assumed there is no fading or shadowing. Then, the MUE HF probability is shown in Fig.~\ref{Fig:No_fading_res}\subref{Fig:MUEHF probability_nofad} for different TTT and $T_{\rm d}$ values. 
We see that as UE velocity increases, 
the MUE HF probability increases. 
On the other hand, 
when the sampling period of L3 filter decreases, 
the MUE HF probability decreases, since the TTT can be initiated earlier. 
For example, the MUE HF probability for UE velocity $80$ km/hr improves by approximately $10$ percent when sampling period is reduced from $150$~ms to $50$~ms. By reducing TTT to $160$~ms we can see that the MUE HF probability becomes almost negligible. This is because the UEs finalize the handover in a quicker way, when compared to larger TTT values.


PUE HF for different TTT and $T_{\rm d}$ values are shown in Fig.~\ref{Fig:No_fading_res}\subref{Fig:PUEHF probability_nofad}.  
We can see that as UE velocity increases, PUE HF probability increases and it is improved when sampling period of L1 and L3 filter decreases. For example, the PUE HF probability for UE velocity $120$ km/hr improves by approximately $5$ percent when sampling period is reduced from $150$~ms to $50$~ms. By reducing the TTT to $160$~ms we can see that there will be no PUE HF probability when no fast fading and shadowing are considered in analysis/simulations.


\begin{figure}[h]
\begin{center}
\includegraphics[width=3.5in]{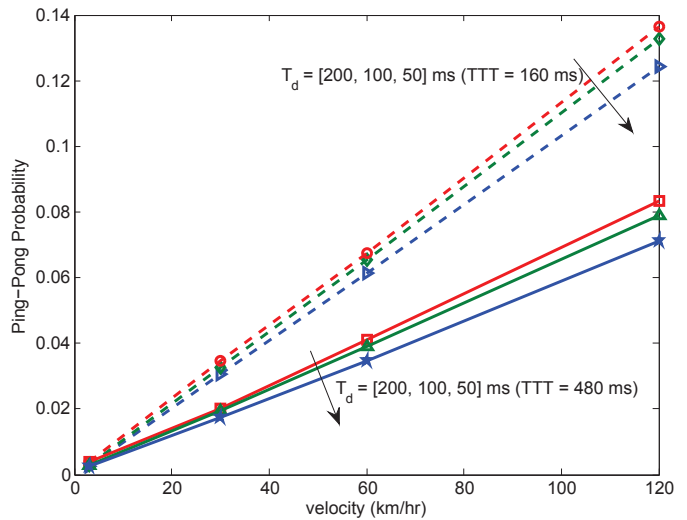}
\captionof{figure}{Simulation plots for ping-pong probabilities as a function of UE velocity for $R=64$~m, $r_{\rm m}=50$~m and $r_{\rm p}=78$~m.}
\label{Fig:PP_probability_nofad}
\end{center}
\end{figure} 


The downside of reducing the sampling period of L1/L3 filter is that it increases the unnecessary handovers called ping-pongs shown in Fig.~\ref{Fig:PP_probability_nofad}. We can see that ping-pong probability increases when the sampling period of L1/L3 filter is reduced from $200$~ms to $50$~ms. The reason for this is that for a larger sampling period of L1/L3 filter, a UE will stay with its serving cell for a longer time before initiating the handover process, which will naturally reduce ping-pong handovers.

%
\subsection{Results with Fading}

\begin{figure}[h]
\begin{center}
\subfloat[MUE HF probability with fast-fading and shadowing (Case~3).]{\includegraphics[width=3.25in]{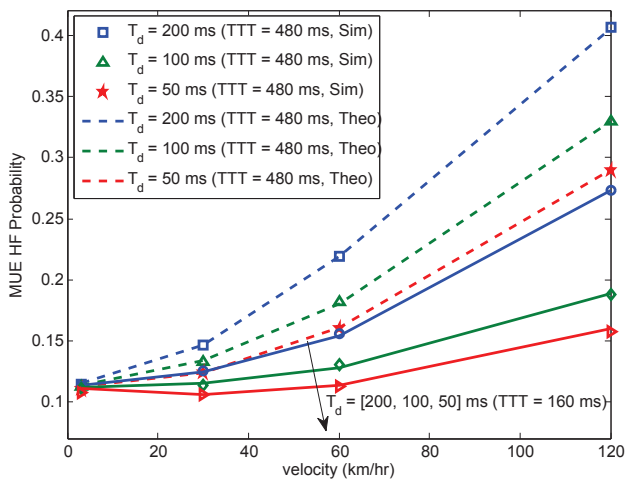}
\label{Fig:MUEHF probability_fad}} \\
\subfloat[PUE HF probability with fast-fading and shadowing (Case~3).]{\includegraphics[width=3.25in]{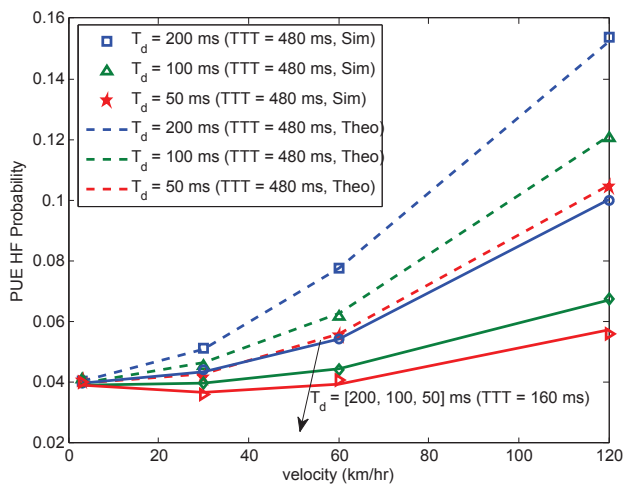}
\label{Fig:PUEHF probability_fad}}
\captionof{figure}{Theoretical (lines) and simulated (markers) 	                                  no fading results as a function of UE velocity for $R=64$~m, $r_{\rm m}=50$~m and $r_{\rm p}=78$~m.}
\label{Fig:Fading_res}
\end{center}
\end{figure}
\vspace{-1mm}
We use the histogram data shown in Fig.~\ref{Fig:histograms_yy_yn} to generate samples for $\hat{r}_{\rm d}$ in a fast-fading and shadowing scenario, 
follow the other simulation assumptions stated previously,
and plot the MUE HF and PUE HF probabilities as a function of UE velocity. MUE HF probability plots are shown in Fig.~\ref{Fig:Fading_res}\subref{Fig:MUEHF probability_fad}. 
We can see that MUE handover performance is degraded for all UE velocities compared to no-fading scenario in Fig.~\ref{Fig:No_fading_res}\subref{Fig:MUEHF probability_nofad}, 
and it is improved when the sampling period ($T_{\rm d}$) of the filter is decreased. We can see that for MUE traveling with a velocity $120$ km/h, the MUE HF is improved by $12.19$ percent when sampling period of the filter is reduced from $200$~ms to $50$~ms.  Results show that even at low UE velocities, there may be on the order of 10\% HF probability. Note that these results consider a worst-case simulation scenario, in which the UE starts its path at the coverage area of a picocell base station as shown in Fig.~\ref{Fig:GeometricModel}.

On the other hand, the PUE HF plots are shown in Fig.~\ref{Fig:Fading_res}\subref{Fig:PUEHF probability_fad}. We notice that the PUE HF probability increases as UE velocity increases and it is improved when the sampling period $T_{\rm d}$ of the filter is decreased. The PUE HF probability 
improves by $3$ percent for PUE traveling with a velocity $120$ km/h in the fading channel conditions. This is due to higher HF probabilities of MUEs at higher speeds.

\begin{figure}[h]
\begin{center}
\includegraphics[width=3.5in]{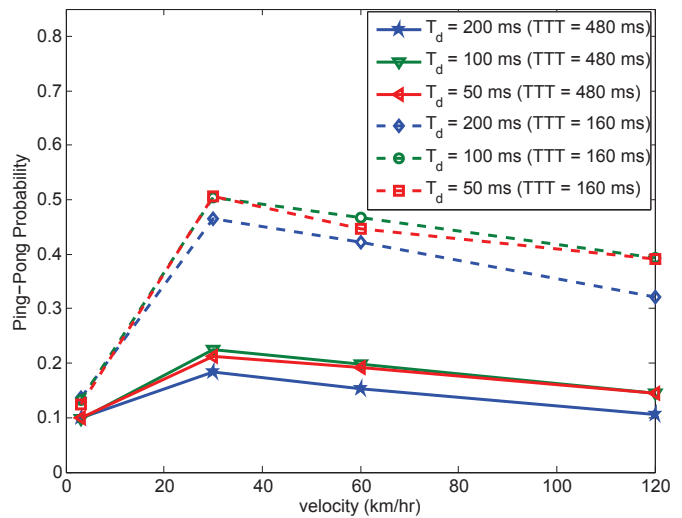}
\captionof{figure}{Simulation plots for fading ping-pong probabilities as a function of UE velocity for $R=64$~m, $r_{\rm m}=50$~m and $r_{\rm p}=78$~m.}
\label{Fig:PP_probability_fad}
\end{center}
\end{figure}

In order to investigate the impact of TTT and $T_{\rm d}$ on ping-ping handover performance, ping-pong probability plots in the fading scenario are shown in Fig.~\ref{Fig:PP_probability_fad}. We can see that there are more ping-pongs when $T_{\rm d}$ and TTT are reduced and this is because, using larger TTT and $T_{\rm d}$, UE will take more time to complete the handover process, as a result ping-pongs is reduced. In the Fig.~\ref{Fig:PP_probability_fad} we can see that there are more ping-pongs in fading scenario compared to no-fading scenario shown in Fig. \ref{Fig:PP_probability_nofad}. 
This is because, in the case of fading scenario and assuming from the picocell perspective, the the link quality of the picocell might decrease below the macrocell because of the channel imperfections. As a result there might be handovers back and forth and UEs time-of-stay will be less than the ping-pong threshold causing more ping-pongs.
\vspace{-2mm}
\section{Conclusion}

In this paper, we developed a simple geometric abstraction to derive analytic and semi-analytic expressions for MUE and PUE HF probabilities for scenarios with no-fading and fading, respectively. Despite the several simplifying assumptions considered in the geometric model and framework, our findings still provide several useful insights. In particular, how different parameters explicitly affect the handover failure of UEs can be captured explicitly, which is not possible in the case of earlier simulation studies available in the literature. The results show that fading degrades the handover performance for all UE velocities. 
The handover performance for both MUEs and PUEs are improved when filtering sampling period is reduced. 
This improvement of handover failure performance is larger in the fading scenario.
In particular, the handover failure performance improved by $12.16$ percent for MUEs in the fading scenario, compared with the $3.92$ percent improvement in the no fading scenario. 
The improvement in the handover failure performance due to shorter L3 sampling period comes at the cost of increased number of ping pong handovers, which will be studied further in our future work.


%

%

\section*{Acknowledgment}
This research was supported in part by the U.S. National Science Foundation under the Grant CNS-1406968, and by the 2014 Ralph E. Powe Junior Faculty Enhancement Award. The authors would like to thank to Farshad Koohifar for reviewing the paper and providing feedback.

\ifCLASSOPTIONcaptionsoff
  \newpage
\fi

\bibliography{References2014}
\bibliographystyle{IEEETran}

\end{document}